\documentclass[sigconf,natbib=true,anonymous=false]{acmart}
\AtBeginDocument{%
  \providecommand\BibTeX{{%
    \normalfont B\kern-0.5em{\scshape i\kern-0.25em b}\kern-0.8em\TeX}}}
\usepackage{url}
\usepackage{graphicx}
\usepackage{amsmath}
\usepackage{amsthm}
\usepackage{booktabs}
\usepackage{algorithm}
\usepackage{algorithmic}
\usepackage{color}
\usepackage{amsfonts}
\usepackage{multirow}
\usepackage{bbding}
\usepackage{subcaption}
\usepackage{pgfplots}
\usepgfplotslibrary{fillbetween}
\usepackage{xcolor}         % colors

\usepackage{tikz}
\usepackage{pgfplots}
\definecolor{mycolor1}{RGB}{90, 51, 180}
\definecolor{mycolor2}{RGB}{120, 81, 151}
\definecolor{mycolor3}{RGB}{84, 144, 196}
\definecolor{mycolor4}{RGB}{205, 35, 33}

\pgfplotsset{linestyle/.style={%
        font=\rmfamily\Labelsize,
        width=0.29\textwidth,
        mark size=1.1pt,
        ylabel near ticks,
        xlabel near ticks,
        xtick=data,
        label style = {font=\scriptsize},
        tick label style = {font=\scriptsize, yshift=0.5ex},
        ylabel shift = -6 pt, 
        xlabel shift = -5 pt,
        title style={yshift=-1.2ex,font=\footnotesize},
        legend image post style={scale=0.5},
        every axis plot/.append style={semithick},
        legend style={font=\scriptsize, mark size=4pt},
        legend columns=1, legend style={/tikz/column 2/.style={column sep=0.5pt}},
        mark options={scale=1}}}

\pgfplotsset{barstyle/.style={%
    ybar,
    ylabel near ticks,
    xlabel near ticks,
    height=4.1cm,
    width=0.29\textwidth,
    label style={font=\scriptsize},
    tick label style={font=\tiny, yshift=0.6ex},
    title style={yshift=-1.2ex,font=\footnotesize},
    ylabel shift = -5 pt, 
    xlabel shift = -2 pt,
    tickwidth         = 3pt,
    xtick=data,
    legend style={font=\tiny, mark size=1pt},
    legend columns=1, legend style={/tikz/column 1/.style={column sep=0.1pt}},
}}

\pgfplotsset{boxstyle/.style={
    boxplot/draw direction=y,
    ylabel near ticks,
    xlabel near ticks,
    height=3.5cm,
    width=0.30\textwidth,
    cycle list={{mycolor1},{mycolor3}},
    label style={font=\scriptsize},
    tick label style={font=\tiny, yshift=0.6ex},
    xtick={1,2},
    ticklabel style={yshift=-0.5ex},
    title style={yshift=-1.2ex,font=\footnotesize},
    ylabel shift = -5 pt, 
    xlabel shift = -2 pt,
    legend style={font=\tiny, mark size=1pt},
    legend columns=1, legend style={/tikz/column 1/.style={column sep=0.1pt}},
}}

\pgfplotsset{xbarstyle/.style={%
    xbar,
    width=0.225\textwidth,
    label style={font=\small},
    tick label style={font=\tiny},
    axis x line*=bottom,
    axis y line*=none,
    height=3.7cm,
    title style={yshift=-0.5ex,font=\small},
    ylabel shift = -5 pt, 
    xlabel shift = -5 pt,
    ylabel near ticks,
    xlabel near ticks,
    tickwidth         = 3pt,
    ytick=data,
    bar width=1mm,
    xlabel style={font=\footnotesize}
}}

\author{
Wei Jiang,
Zhongkai Yi\footnote{corresponding author},
Li Wang,
Hanwei Zhang,
Jihai Zhang,
Fangquan Lin,
Cheng Yang}
\affiliation{
  \city{Alibaba Group, Hangzhou}
  \country{China}}
\renewcommand{\shortauthors}{Wei Jiang et al.}

\begin{document}

% \maketitle
\title{A Stochastic Online Forecast-and-Optimize Framework for Real-Time Energy Dispatch in Virtual Power Plants under Uncertainty}

% \author{Anonymous}

\begin{abstract}
Aggregating distributed energy resources in power systems significantly increases uncertainties, in particular caused by the fluctuation of renewable energy generation. This issue has driven the necessity of widely exploiting advanced predictive control techniques under uncertainty to ensure long-term economics and decarbonization. In this paper, we propose a real-time uncertainty-aware energy dispatch framework, which is composed of two key elements: \textit{(i)} A hybrid forecast-and-optimize sequential task, integrating deep learning-based forecasting and stochastic optimization, where these two stages are connected by the uncertainty estimation at multiple temporal resolutions; \textit{(ii)} An efficient online data augmentation scheme, jointly involving model pre-training and online fine-tuning stages. In this way, the proposed framework is capable to rapidly adapt to the real-time data distribution, as well as to target on uncertainties caused by data drift, model discrepancy and environment perturbations in the control process, and finally to realize an optimal and robust dispatch solution. The proposed framework won
the championship in CityLearn Challenge 2022, which provided an influential opportunity to investigate the potential of AI application in the energy domain. In addition, comprehensive experiments are conducted  to interpret its effectiveness in the real-life scenario of smart building energy management. 
\end{abstract}

\maketitle

\section{Introduction}\label{sec:intro}

In recent years, virtual power plants (VPPs) as an aggregation of distributed energy resources (DERs), have enabled the efficient exploitation of a large number of renewable energy into power systems~\cite{nosratabadi2017comprehensive}.  While the sustainable development has been promoted to a new level, however, the intermittent and fluctuating nature of renewable energy generation also poses a huge challenge to the traditional dispatching mode~\cite{olivares2014trends}. In fact, the effectiveness of dispatch largely relies on the precise prediction of future information \cite{camacho2013model}. To guarantee the economy and decarbonization of long-term operations, it is necessary to develop an uncertainty-aware decision-making method for VPPs.
For further investigation, we first classify the uncertainty factors into three types:
\begin{enumerate}
    \item \textit{Data drift:}  %The relation between input data and the target variables changes over time
    The data distribution changes over time
    \cite{gama2014survey}. %For example, the sequential transition in time-series of renewable energy generation can be fluctuated.
    Apart from the intermittency and fluctuation of the renewable energy generation (\textit{e.g.}, wind power and solar power), time-varying electricity prices and flexible load demand can further lead to inferior dispatch solution~\cite{yu2019uncertainties}. 
    \item \textit{Model discrepancy:} 
    The model assumption does not fully correspond to the real mechanism of data generation, which can result in a large variance in the dispatch effect \cite{arendt2012quantification}. %in the control process. 
    For instance, prediction errors appear ubiquitously and show limited generalization ability, due to structural mismatch between the forecasting model and the real process. %, \textit{e.g.}, in the form of an empirical data fit with deep neural network.
    \item \textit{Environment perturbation:} 
    The environment simulator does not match the true process that occurs. For example, the physical system of energy storage devices may involve a mismatch of parameters \cite{wang2021adaptive}. As a result, random perturbation widely occurs in the control process.
    % ($<-$ simulator: feedback correction; )
\end{enumerate}

Existing approaches, such as conventional model predictive control (MPC) framework, have employed rolling-horizon control to correct the parameter mismatch in environment with real-time feedback mechanism \cite{camacho2013model,hewing2020learning}. However, the other forms of uncertainty still weaken the advantages of VPPs. As a representative example of industrial applications, the sequential MPC framework can usually be decomposed into point forecasting of the target variables (\textit{e.g.}, solar power, load demand, \textit{etc.}), followed by deterministic optimization, which cannot capture the uncertainties of the probabilistic data distribution \cite{muralitharan2018neural,elmachtoub2022smart}. 
More recently, stochastic approaches have received more attention which eliminate the influence of some uncertain factors \cite{yu2019uncertainties,kong2020robust}. Despite these recent advances, the increased model complexity with numerous DERs makes it difficult to meet real-time dispatch requirements; at the same time, the existing schemes are not designed for real scenarios.

Motivated by these requirements in practice, we aim to provide decision-making support on real-time dispatch commands under uncertainty, to achieve the economy and decarbonization of long-term operations of VPPs. %, and to promote the maximum consumption of renewable energy.
The main contributions are as follows:
\begin{itemize}
    \item A novel and complete framework---\textbf{S}tochastic 
 \textbf{O}nline \textbf{F}orecast-and-\textbf{O}ptimize Framework (SOFO) is proposed for real-time uncertainty-aware decision-making. To the best of our knowledge, this is the first work to comprehensively investigate various types of uncertainties in the control process, from the perspectives of both data drift and model discrepancy.
    \item To tackle the explored uncertainties, two data-driven modules are proposed to achieve the optimal and robust dispatch control, including a hybrid forecast-and-optimize sequential task integrating deep learning-based forecasting and stochastic optimization, and an efficient online data augmentation scheme involving model pre-training and fine-tuning stages.
   \item  The proposed framework is application-driven, which won 
the 1st place of the CityLearn Challenge 2022 
\cite{T8/0YLJ6Q_2023}, which provided an influential opportunity to investigate the potential of AI for application in the energy domain.
   A novel and practical industrial application is illustrated in a real-life scenario of smart building energy management. Empirical comparison against various baselines is demonstrated.
\end{itemize}

% in a complex environment of distributed energy sources, users and energy storage devices. 
% realizes fast decision-making on real-time dispatching commands for virtual power plants, and promotes economic and environmentally friendly operation of virtual power plants. 
% In addition, in the state of real-time data flow, based on the predictive control with feedback correction, the scheme proposes online data augmentation to achieve rapid update of model parameters adapted to real-time data by combining deep learning and reinforcement learning for pre-training and fine-tuning to improve model accuracy and cope with uncertainties caused by model mismatch, time-varying and random perturbations in the control process to realize closed-loop energy control of the system.
% it copes with uncertainties caused by model mismatch, temporal variations and random perturbations in the control process,

\section{Related Work} \label{sec:relatedwork}
% \paragraph{Title.} text.
% ~\cite{gottlob:nonmon}
% ~\shortcite{nebel:jair-2000}
In this section, we briefly review the classic approaches of model predictive control, and then focus on the recent advances of energy dispatch. Specifically, we discuss the two essential branches of dispatch approaches in VPPs, including uncertainty-aware control and learning-based control.
\subsection{Model Predictive Control}
As a conventional decision-making strategy, the MPC framework has been widely studied and employed in many fields of application \cite{wang2009fast,berberich2020data}, including the power systems for energy dispatch \cite{rawlings2000tutorial,kouro2008model,ernst2008reinforcement,sultana2017review}. Generally speaking, MPC is a constrained optimal control strategy, based on finite-horizon iterative optimization of a plant model. %Taking into account the future predictive information, it optimizes a finite temporal horizon, but implements only the current timestamps, then optimizes again, iteratively. 
%As a result, both the prediction and control paths are updated continuously.
Recently, one center of application has been specifically switched to VPPs, for example, energy management of smart buildings
\cite{drgovna2020all,arroyo2022reinforced}. The task can be decomposed into forecasting and optimization stages. Usually, the target variables of forecasting consist of power generation, load demand, market price, \textit{etc.} Regarding optimization, cases with multiple objectives are quite common, including electricity price cost and greenhouse gas emission. Dispatch solutions for energy devices in VPPs are optimized with constraints considering power balance and device attributes \cite{yu2019uncertainties,drgovna2020all}.
\subsection{Uncertainty-Aware Control}
In real-life applications, implementations of dispatch strategies suffer from multiple uncertainty factors.
As pointed out in \cite{lauro2014adaptive}, considering
forecast uncertainty can lead to increased energy savings in the range of 15\% to 30\% for VPPs. Existing works that mitigate uncertainties include stochastic approaches and adaptive approaches \cite{heirung2018stochastic}. Firstly, stochastic approaches 
include chance-constraint approximations and feedback parameterizations, which both require prior knowledge of the system uncertainties \cite{yan2018stochastic,paulson2020stochastic}. Another representative stochastic approach is scenario-based, generating a number of realizations of the stochastic variables \cite{shang2019data,bradford2020stochastic}. Alternatively, adaptive approaches mainly focus on the online update of the predictive control \cite{ioannou2012robust,aastrom2013adaptive}. For example, predictions and dispatch solutions can be updated via the autoregression in a rolling-horizon framework \cite{liu2018model}.
Adaptive control approaches are typically restricted to specific model types, which affects the ability to generalize.
In recent years, adaptive approaches have been revised in order to combine various machine learning approaches, as discussed in the following.

\subsection{Learning-Based Control}
As an active area that involves new applications, 
learning-based control aims to learn the prediction model from data with uncertainty quantification \cite{hewing2020learning}. Compared to typical adaptive control, %that is restricted to specific model types, 
learning-based control involves various statistical learning methods to improve system dynamics in various domains \cite{bujarbaruah2018adaptive,jiang2021dynamic,lin2021dynamic}. In the literature on VPPs, there is a multitude of data-driven approaches to represent the prediction variables %which are incapable of dealing with various types of uncertainties 
\cite{hernandez2013multi,tascikaraoglu2014adaptive,macdougall2016applying,gougheri2021optimal}. Common prediction models consist of tree-based ones, including random forests and boosting trees \cite{jain2017data,smarra2018data}. Recently, deep learning-based methods have been adopted to provide 
more accurate prediction results \cite{gougheri2021optimal}.
One essential challenge is to overcome the  overfitting, caused by the differences in the data distribution during training and inference stages.

In this paper, we combine both the scopes of stochastic and adaptive approaches, to target various types of uncertainty simultaneously. First, the prediction model and scenario generation are connected by the uncertainty estimation at multiple temporal resolutions. In addition, the proposed prediction model is learning-based, which enables the two-stage adaptive approach of model pre-training and fine-tuning. Specifically, application-driven method in the field of VPPs is introduced in the following section.

\section{Problem Statement and Definitions} \label{sec:problem}
%In this work, our aim is to provide energy management solutions for VPPs to achieve the economy and decarbonization of long-term operations. 
Formally, we denote that a VPP consists of various DERs, including a set of power generation devices $\mathcal{G}$, a set of user devices $\mathcal{U}$ and a set of energy storage devices $\mathcal{S}$. The dispatch period with total $T$ timestamps. Given a timestamp $t \in \mathcal{T}=\{1, \ldots, T\}$, %we define the maximum value of power generation of device $g \in \mathcal{G}$ as $P_{g,t}^{\rm max}$ and 
we define the load demand of the user $u \in \mathcal{U}$ as $L_{u,t}$. 
Meanwhile, we assume the market price unit at a timestamp $t$ is $p_{t}$. More generally, if there are a set of multiple market $\mathcal{M}$ (\textit{e.g.}, electricity market and carbon market), then $p_{t}$ is the average of the price units among all markets.

The target variables of energy management include the electricity consumption from grid $P_{\text{grid},t}$, the power generation $P_{g,t}$ of device $g \in \mathcal{G}$, the charging or discharging power $P^+_{s,t}$ or $P^-_{s,t}$ and the state-of-charge $E_{s,t}$ of device $s \in \mathcal{S}$. 
% We aim to decide the  variables of energy management, including the charging variable from the power generation device $g$ to the energy storage device $s \in \mathcal{S}$ (denoted by  $x^+_{g \rightarrow s,t}$), the discharging variable from $s$ to the user $u$ (denoted by $x^-_{s \rightarrow u,t}$) and  directly usage of power generated from $g$ to user $u$ (denoted by $x_{g \rightarrow u,t}$). 
Let the set of decision variables be $X = \{P_{\text{grid},t},  P_{g,t}, P^+_{s,t}, P^-_{s,t}, E_{s,t} \}$, where $t \in \mathcal{T}$, $s \in \mathcal{S}$, $g\in \mathcal{G}$, %$X = \{x^+_{g \rightarrow s,t}, x^-_{s \rightarrow u,t}, x_{g \rightarrow u,t} \}$, %where ${t\in \{1, \ldots,T\}$, $s \in \mathcal{S}$, $g\in \mathcal{G}, u \in \mathcal{U}}$, 
then the objective is to achieve the economic optimum of all markets on average, formally defined as follows: 

% \begin{subequations}\label{eq:obj_problem}
% % \begin{gather}
% \begin{align}
% \text{s.t.:} \quad &  P_{g,t} \le P_{g,t}^{\rm max} & \pushright{\hfill g \in \mathcal{G}, t \in \{1, \ldots,T \}} \label{eq:cons1} \\
% &   0 \le P^+_{s,t} \le {P^+_{s,t}}^{\rm max}  &\pushright{\hfill s \in \mathcal{S}, t \in \{1, \ldots,T \}} \label{eq:cons2} \\
% &   0 \le P^-_{s,t} \le {P^-_{s,t}}^{\rm max}  &\pushright{\hfill s \in \mathcal{S}, t \in \{1, \ldots,T \}} \label{eq:cons3} \\ 
% & P^+_{s,t} \cdot  P^-_{s,t}  = 0 & \pushright{\hfill s \in \mathcal{S}, t \in \{1, \ldots,T \}} \label{eq:cons4} \\ 
% \end{align}
% \end{subequations}
\begin{subequations}\label{eq:obj_problem}
\begin{equation}
\mathop{\mathrm{minimize}}_{X} \quad \sum_{t=1}^T p_t \cdot P_{\text{grid},t}
 \tag{\ref{eq:obj_problem}}
\end{equation}
% \begin{gather}
\text{s.t.:} 
\begin{align}
&  P_{\text{grid},t} \ge 0   & {\hfill  t \in \mathcal{T}} \label{eq:cons0} \\
&  P_{g,t}^{\rm min} \le P_{g,t} \le P_{g,t}^{\rm max}& {\hfill \quad g \in \mathcal{G},\, t \in \mathcal{T}} \label{eq:cons1} 
\end{align}
\begin{equation}
\left. \begin{aligned}
&   0 \le P^+_{s,t} \le {P^+_{s,t}}^{\rm max}  \\ %\pushright{\hfill s \in \mathcal{S}, t \in \{1, \ldots,T \}} \label{eq:cons2} \\
&   0 \le P^-_{s,t} \le {P^-_{s,t}}^{\rm max}  \\ 
& P^+_{s,t} \cdot  P^-_{s,t}  = 0 %\pushright{\hfill s \in \mathcal{S}, t \in \{1, \ldots,T \}} \label{eq:cons4} \\ 
\end{aligned} \qquad \qquad \quad \, \right\}  s \in \mathcal{S},\, t \in \mathcal{T} \label{eq:cons2} 
\end{equation}
\begin{equation}
\begin{aligned}
&  E_{s,t}^{\rm min} \le E_{s,t} \le E_{s,t}^{\rm max} & {\hfill s \in \mathcal{S}, t \in \mathcal{T}}\\
& E_{s,t} = E_{s,t-1} + P^+_{s,t} - P^-_{s,t} & {\hfill s \in \mathcal{S}, t \in \mathcal{T} \setminus \{1\}}
\end{aligned} %\right\} & {\hfill s \in \mathcal{S}, t \in \{1, \ldots,T \}} \label{eq:cons3} 
\label{eq:cons3}
\end{equation}
\begin{equation}
 % \resizebox{.99\linewidth}{!}{
P_{\text{grid},t} + \sum_{g \in \mathcal{G}} P_{g,t} + \sum_{s \in \mathcal{S}} P^-_{s,t}  = \sum_{s \in \mathcal{S}} P^+_{s,t} + \sum_{u \in \mathcal{U}} L_{u,t}  \quad  t \in \mathcal{T}  \label{eq:cons4}
% }
\end{equation}
\end{subequations}
%To facilitate the understanding of the above constraints, 
Their brief explanation of the above constraints is provided as follows with the necessary description of parameters:
\begin{itemize}
    \item[(\ref{eq:cons0})] Bounds of electricity consumption from grid;%: larger than zero and without upper bounds;
    \item[(\ref{eq:cons1})] Power generation device attributes: lower bounds ($P_{g,t}^{\rm min}$) and upper bounds ($P_{g,t}^{\rm max}$) of power;
    \item[(\ref{eq:cons2})] Energy storage attributes: upper bounds of charging (${P^+_{s,t}}^{\rm max}$), upper bounds of discharging (${P^-_{s,t}}^{\rm max}$) and status constraints; %(either charging or discharging at each timestamps);
    \item[(\ref{eq:cons3})] State-of-charge constraints:  lower bounds ($E_{s,t}^{\rm min}$), upper bounds ($E_{s,t}^{\rm max}$) and update formulas;
    \item[(\ref{eq:cons4})] Power balance constraints of the VPP system.
\end{itemize}

% \subsection{Energy Storage Devices Models}
% \subsection{Multi-objective Task}
However, in practice, when planning dispatch, we are unable to obtain the precise values of load, power generation, and market price in advance. Therefore, we propose a stochastic online forecast-and-optimize framework, as detailed in the following section.

\begin{figure*}[t]
\centering
\includegraphics[width=0.9\textwidth]{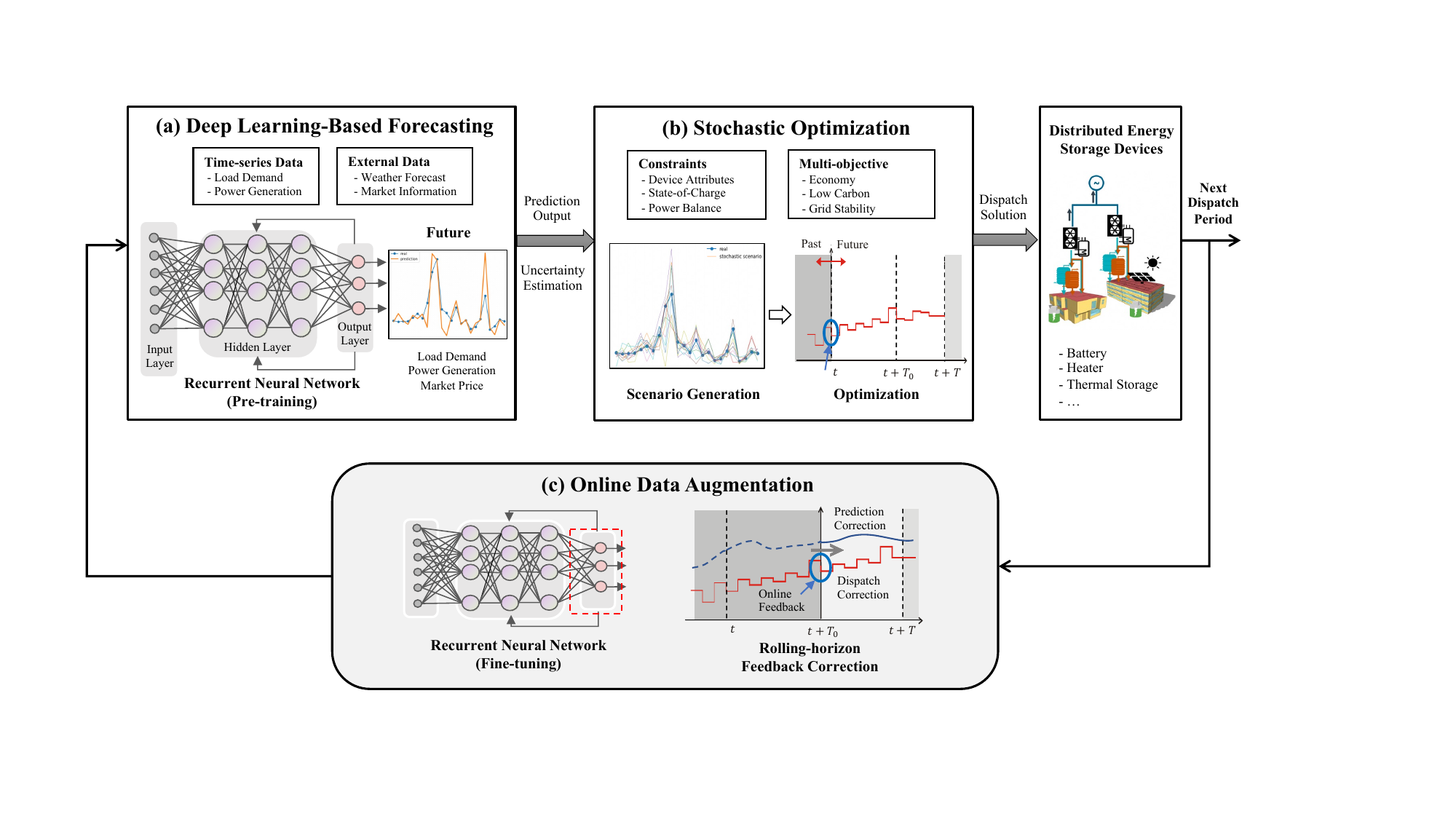}
\caption{ An Overview of the SOFO framework. It consists of two major modules: \textit{i)} A hybrid forecast-and-optimize sequential task, as shown in the subplot (a) deep learning-based forecasting, followed by the subplot (b) stochastic optimization; \textit{ii)} An online data augmentation scheme as demonstrated in the subplot (c). Based on these stochastic and adaptive techniques, the SOFO framework contributes to generate optimal and robust dispatch solutions under uncertainties.
}
% It is a symmetric structure on the user side and the item side.
% The middle part (the black arrows) represents the recommendation prediction (RP) module (Section \ref{sec:rpm}). It generates the intrinsic and extrinsic factor representations ($\Gin$ and $\Gex$) for producing the recommendation prediction $y^{\prime}$. 
% The side parts are two contrastive intrinsic-extrinsic disentanglement (CIED) modules. Each CIED includes a context-invariant contrastive learning component (the red arrows, Section \ref{sec:method_CICL}), and a disentangling component (the blue arrows, Section \ref{sec:method_biDis}) to ensure the success of the factor learning.
% The losses generated through these modules ($\mathcal{L}_{RP},\mathcal{L}_{CICL},\mathcal{L}_{bi\text{-}appr},\mathcal{L}_{Dis}$) will be optimized as a two-step multi-task training (Section \ref{sec:training}).}
\label{fig:framework}
\end{figure*} 

\section{Stochastic Online Forecast-and-Optimize Framework}
The overview of the proposed framework is visualized in Figure \ref{fig:framework}, which consists of two modules: \textit{(i)} A hybrid forecast-and-optimize sequential task, integrating deep learning-based forecasting and stochastic optimization, where these two stages are connected by the uncertainty estimation at multiple temporal resolutions; %A recommendation prediction (RP) module that takes a user and an item as input, and combines them with a set of contexts, to generate intrinsic and extrinsic factor representations for both the user and the item. The predicted probability $y^{\prime}$ is then jointly learned from these representations.
\textit{(ii)}An efficient online data augmentation scheme, involving model pre-training and online fine-tuning stages. %A contrastive intrinsic-extrinsic disentangled (CIED) module is applied to both the user and the item sides to support the intrinsic and extrinsic factor learning. The module contains a context-invariant contrastive learning component and a disentangling component, to ensure the learned factors satisfy Definition \ref{def:in_ex_factor}.

\subsection{Stochastic Forecast-and-Optimize}
\paragraph{Deep Learning-based Forecasting Model.}
Firstly, we build forecasting models to estimate the variables at each timestamp $t$ of the control horizon, as illustrated in subplot (a) of Figure \ref{fig:framework}. The target variables consist of: \textit{i)} Capacity of 
renewable power generation $P_{g,t}^{\rm max}$; \textit{ii)} Load demand $L_{u,t}$; \textit{iii)} Market price $p_{t}$. 
Note that the other bounding parameters in the problem (\ref{eq:obj_problem}) are constant given the physical models of the devices. The input features of the forecasting models can be classified into the following genres: \textit{i)} Discrete time-related features, such as hours of day, days of week, months of year, \textit{etc.}; \textit{ii)} Historical sequences of target variables; \textit{iii)} Exogenous features, such as weather predictions and market information. 

Deep learning-based models are involved in the forecasting task. Here we mainly describe the recurrent neural network (RNN) %with a gated recurrent Unit (GRU) 
as a representative algorithm, but note that the proposed framework can be naturally superimposed on any deep learning-based algorithm (\textit{e.g.}, CNN, transformer). %GRU is one of the most commonly-applied types of recurrent neural network with a gating mechanism \cite{cho-etal-2014-properties}. 
RNN has shown good performance in many sequential tasks of signal processing \cite{athiwaratkun2017malware,10.1016/j.neucom.2019.04.044}. %and natural language processing 
Though less active in the energy domain, there are studies showing that RNN is well-performed in smaller-sample datasets, %compared to other variants of recurrent network 
especially with specific design of gating mechanisms \cite{chung2014empirical}. %,gruber2020gru}. 
% More formally, at time $t$, given an element $x_t$ in the input sequence, GRU layer updates the hidden state from $h_{t-1}$ to $h_t$ as follows:
% \begin{equation*} \label{eq:gru}
% \begin{split}
% r_{t}= &\sigma\left(W_{i r} x_{t}+b_{i r}+W_{h r} h_{(t-1)}+b_{h r}\right)\,, \\
% z_{t}= & \sigma\left(W_{i z} x_{t}+b_{i z}+W_{h z} h_{(t-1)}+b_{h z}\right)\,, \\
% n_{t}= & \tanh \left(W_{i n} x_{t}+b_{i n}+r_{t} \circ \left(W_{h n} h_{(t-1)}+b_{h n}\right)\right)\,, \\
% h_{t}= &\left(1-z_{t}\right) \circ n_{t}+z_{t} \circ h_{(t-1)}\,,
% \end{split}
% \end{equation*}
% where $r_t$, $z_t$, and $n_t$ represent the reset, update and new gates, respectively; $\sigma$ is the sigmoid function and $\circ$ denote the Hadamard product.
Formally, given a sequence of input $x = (x_1, x_2, \ldots, x_T)$, the RNN layer expresses its recurrent hidden state $h_t$ and the output variable $y_t$ by:
\begin{equation*}
% \begin{split}
h_t = \phi_1 \left(h_{t-1}, x_t\right)\,, \quad y_t = \phi_2 \left(h_t\right)\,, \quad  t \in \mathcal{T} \,.
% \end{split}
\end{equation*}
where $\phi_1$ and $\phi_2$ are non-linear functions such as the composition of an activate function with affine transformations.
% More generally we define the neural network as a function $f$, 
%  and the predicted output at timestamps $t$ as $\hat{y}_t = f(x_1, \ldots, x_t)$ then the loss function can be defined as: 
% \begin{equation*}
%     \mathcal{L}(f(x), y) = \sum_{t=1}^T \mathcal{L}(\hat{y}_t, y_t) \,.
% \end{equation*}
% For each target variable, we build an RNN model. % based on their time-series sequences, as well as the exogenous features. 
By maximizing the likelihood of the models given the training data, we denote the learned models as $f_{P_g}$, $ f_{L_u}$, $f_{p}$ for capacity of power generation, load demand and market price, respectively.

\paragraph{Stochastic Optimization.}
Then we apply the trained models to infer the predictions of the target variables on the control horizon: $\hat{P}_{g,t}^{\rm max}$, $\hat{L}_{u,t}$ and $\hat{p}_t$, where $t \in \mathcal{T}$. In addition, the variance of forecasting models are estimated on the validation data by the empirical prediction errors, denoted by $\hat{\Sigma}_{P_g}$, $\hat{\Sigma}_{L_u}$ and $\hat{\Sigma}_{p}$, to quantify the level of uncertainty.
To further address the uncertainty of the forecasting, we propose stochastic optimization for real-time energy dispatch, as demonstrated in subplot (b) of Figure \ref{fig:framework}. 
Realizations of the stochastic Gaussian process are generated with mean and variance parameters equal to ($\hat{P}_{g,t}^{\rm max}$, $\hat{\Sigma}_{P_g}$), ($\hat{L}_{u,t}$, $\hat{\Sigma}_{L_u}$) and ($\hat{p}_t$, $\hat{\Sigma}_{p}$), respectively. That means,
Gaussian noises are added on the predicted values, to generate a multi-scenarios optimization problem which we solve coordinately. Consider the case with $N$ scenarios, then the $n$-th realizations of the parameters can be denoted by ($n \in \mathcal{S_N}$): 
\begin{equation*}
    \begin{split}
        (\tilde{P}_{g}^{\rm max})^n &= \left[ (\tilde{P}_{g,1}^{\rm max})^n, (\tilde{P}_{g,2}^{\rm max})^n, \ldots, (\tilde{P}_{g,T}^{\rm max})^n \right]\,, \\
                (\tilde{L}_{u})^n &= \left[ (\tilde{L}_{u,1})^n, (\tilde{L}_{u,2})^n, \ldots, (\tilde{L}_{u,T})^n \right]\,,
        \\
                        (\tilde{p})^n &= \left[ (\tilde{p}_{1})^n, (\tilde{p}_{2})^n, \ldots, (\tilde{p}_{T})^n \right] \,.
    \end{split}
\end{equation*}
Based on the above notation, we solve the stochastic optimization problem by substituting the objective in problem (\ref{eq:obj_problem}) with:
\begin{equation}\label{eq:stoch_opt}
\mathop{\mathrm{minimize}}_{X} \quad \sum_{t=1}^T \mathbb{E}_{n\in \mathcal{S}_{\mathcal{N}}}(\tilde{p}_t)^n \cdot P_{\text{grid},t} \,.
\end{equation}
% \begin{gather}
Meanwhile, the constraint (\ref{eq:cons1}) is replaced by:
\begin{equation*}
{P_{g,t}^{\rm min} \le P_{g,t} \le (\tilde{P}_{g,t}^{\rm max})^n } \quad {n \in \mathcal{S_N}, g \in \mathcal{G}, t \in \mathcal{T}} \,.
 \end{equation*}
 And the constraint (\ref{eq:cons4}) is replaced by:
\begin{equation*}
\resizebox{.99\linewidth}{!}{$
            \displaystyle
P_{\text{grid},t} + \sum_{g \in \mathcal{G}} P_{g,t} + \sum_{s \in \mathcal{S}} P^-_{s,t}  = \sum_{s \in \mathcal{S}} P^+_{s,t} + \sum_{u \in \mathcal{U}} (\tilde{L}_{u,t})^n  \quad n \in \mathcal{S_N},\,  t \in \mathcal{T} \,.
$}
\end{equation*}
By solving the optimization problem (\ref{eq:stoch_opt}), we obtain the dispatch decision solution: $\dot{X} = \{\dot{P}_{\text{grid},t},  \dot{P}_{g,t}, \dot{P}^+_{s,t}, \dot{P}^-_{s,t}, \dot{E}_{s,t} \}$\,.
 
\subsection{Online Data Augmentation}
As shown in subplot (c) of Figure \ref{fig:framework}, the second module in the proposed SOFO framework is online data augmentation. %to adapt to the issue of data drift. 

% \paragraph{Self-Adaptive Correction}
% \textcolor{red}{
% To take into account the data drift issue, we use a linear prediction adjustment, by comparing the historical real values (\textit{e.g.,} ${L}_{u,t}$) and the predicted values (\textit{e.g.,} $\hat{L}_{u,t}$). %For instance, denote the real values of history load demand of user $u$ at timestep $t$ as ${L}_{u,t}$ and the corresponding prediction as $\hat{L}_{u,t}$. 
% The self-adaptive correction parameter is updated at each timestep $T_{\rm sa}$ as follows (similarly for other parameters):
% \begin{equation*}
% \hat{\beta}_{T_{\rm sa}}=\underset{\beta}{\operatorname{argmin}} \sum_{t=1}^{T_{\rm sa}}\left(\beta \cdot \hat{L}_{u,t}-L_{u,t}\right)^2 = \frac{\sum_{t=1}^{T_{\rm sa}} L_{u,t} \cdot \hat{L}_{u,t}}{\sum_{t=1}^{T_{\rm sa}} {\hat{L}_{u,t}}^2} \,.
% \end{equation*}
% }

\paragraph{Pre-training and Fine-tuning Scheme}
% In practice, the real-time energy dispatch processes as a periodic task (\textit{e.g.}, daily dispatch). 
For neural network-based forecasting models, the online data augmentation can be handled via an efficient scheme involving pre-training and fine-tuning:
\textit{(i)} With offline training data, we train the neural network from scratch $f_{P_g}$, $ f_{L_u}$, $f_{p}$; 
\textit{(ii)} With online data, we fine-tune the neural network, either when the fine-tuning horizon (denoted by $T_{\rm ft}$) is reached, or the prediction errors are larger than a given threshold. Regarding the fine-tuning techniques, we freeze the weights of the first few layers of the pre-trained neural networks, then fine-tune the last layer to specifically adapt to the distribution of the newly collected data. Denote the fine-tuned neural network as $\tilde{f}_{P_g}$, $\tilde{f}_{L_u}$, $\tilde{f}_{p}$.

\paragraph{Rolling-horizon Feedback Correction}
In addition, rolling-horizon control is involved (with rolling horizon denoted by $T_{\rm rl}$). At a timestamp $t$, we predict and optimize for the next $T$
timestamps; then after $T_{\rm rl}$, we take new measurements, and correct the previous predictions and actions. This procedure is conducted repeatedly.

% \subsection{Summary of Framework}
To conclude, the summary of the proposed SOFO framework is described in Algorithm \ref{alg:algorithm}.

\begin{algorithm}[tb]
    \caption{Stochastic Online Forecast-and-Optimize Framework}
    \label{alg:algorithm}
    \begin{flushleft}
            \textbf{Input}: Training data $\mathcal{D}_{\rm train}$, validation data $\mathcal{D}_{\rm val}$.\\
    \textbf{Parameter}: Set of hyperparameters of neural networks; constant values of bounding parameters in problem (\ref{eq:stoch_opt}). 
    % \textbf{Output}: Dispatch solution.
    \end{flushleft}
    \begin{algorithmic}[1] %[1] enables line numbers
        \STATE Pre-train forecasting models with $\mathcal{D}_{\rm train}$ $\rightarrow$ $f_{P_g}$, $ f_{L_u}$, $f_{p}$\,;
        \STATE Estimate prediction variance $\mathcal{D}_{\rm valid}$ $\rightarrow$ $\hat{\Sigma}_{P_g}$, $\hat{\Sigma}_{L_u}$, $\hat{\Sigma}_{p}$. \\
        % \STATE Let $t = 1$.
        \WHILE{$t \in \mathcal{T}$ in control horizon}
        % \STATE $t = t + 1$
        \STATE // \textit{\textbf{ Forecast-and-Optimize}}
        \STATE Infer predictions $\rightarrow$ $\hat{P}_{g,t}^{\rm max}$, $\hat{L}_{u,t}$ and $\hat{p}_t$, $t \in \mathcal{T}$\,;
        \STATE Scenarios generation with
        ($\hat{P}_{g,t}^{\rm max}$, $\hat{\Sigma}_{P_g}$), ($\hat{L}_{u,t}$, $\hat{\Sigma}_{L_u}$), ($\hat{p}_t$, $\hat{\Sigma}_{p}$)\,;
        \STATE Solving stochastic optimization problem (\ref{eq:stoch_opt}) $\rightarrow$ $\dot{X} = \{\dot{P}_{\text{grid},t},  \dot{P}_{g,t}, \dot{P}^+_{s,t}, \dot{P}^-_{s,t}, \dot{E}_{s,t} \}$\,. \\
        \STATE // \textit{\textbf{ Online Data Augmentation }}
        \IF {$t > T_{\rm ft}$ fine-tuning horizon or prediction errors $\ge \varepsilon$}
        \STATE Update online datasets $\tilde{\mathcal{D}}_{\rm train}$ and $\tilde{\mathcal{D}}_{\rm val}$ \,;
        \STATE Fine-tune forecasting models $\tilde{f}_{P_g}$, $ \tilde{f}_{L_u}$, $\tilde{f}_{p}$\,;
        \STATE Estimate prediction variance $\tilde{\mathcal{D}}_{\rm val}$ $\rightarrow$ $\hat{\Sigma}_{P_g}$, $\hat{\Sigma}_{L_u}$, $\hat{\Sigma}_{p}$\,.
        % \ELSE
        % \STATE Perform task B.
        \ENDIF
        \IF {$t > T_{\rm rl}$ in rolling-horizon}
        \STATE Infer predictions $\rightarrow$ $\hat{P}_{g,t}^{\rm max}$, $\hat{L}_{u,t}$  and optimize $\rightarrow$ $\dot{X}$\,.
        % \ELSE
        % \STATE Perform task B.
        \ENDIF
        \ENDWHILE        
        \STATE \textbf{return} Dispatch solution $\dot{X}$\,.
    \end{algorithmic}
\end{algorithm}

\section{Experiment Study}
%\textcolor{red}{TO DISCUSS: shall we take the citylearn dataset as a simulation example, and take another case such as chengtao yuan data as a real-life case study?}
% In this section, we will demonstrate the experiment results.
% We focus on comparing the performance of the SOFO framework against the state-of-the-art methods and demonstrating the effectiveness of each framework component. %; \textit{iii)} The ability to address 
%the uncertainties of both the data and the modeling.
% We aim at answering the following questions: %related to the performance of the proposed framework:\\
% \textbf{\emph{RQ1}}. How does the proposed NASSL perform compared to the state-of-the-art methods of news recommendation? \\
% \textbf{\emph{RQ2}}. How do different components of NASSL benefit its performance, \textit{i.e.}, the effectiveness of the neighbor generator, neighbor-augmented encoder, and self-supervision module?
\subsection{Experiment Setup}
\paragraph{Dataset}
Experiments are conducted on building energy management from the CityLearn Challenge 2022 \cite{T8/0YLJ6Q_2023}. The challenge utilizes 1 year of electricity demand and photo-voltaic generation data from 17 single-family buildings  in Fontana, California. 
In this competition, we won the $1^{\rm st}$ place\footnote{\url{www.aicrowd.com/challenges/neurips-2022-citylearn-challenge/leaderboards}} using the proposed framework for energy management. 
After the competition, we have investigated various extension of the framework modules to further improve the performance. %, as illustrated in Section \ref{sec:abl}.
\begin{table*}[t]
\centering
% \large
\caption{Comparison of dispatch performance in the testing set (a community with 7 buildings) with baselines over a year, as well as overall performance in the entire district (communities with 17 buildings in total). All cost presented above are normalized against the simple baseline without electrical energy storage in batteries, such that lower values of cost are preferred. The \textit{Improv.} row show the relative improvements of SOFO framework over the best performed baselines for each metric, respectively, where the positive value indicates that the cost is relatively reduced.}
\begin{tabular}{lcccccccc}
\toprule

\multirow{2}{*}{\textbf{}}  & \multicolumn{4}{c}{\textbf{Overall Performance}} & \multicolumn{4}{c}{\textbf{Testing Set Performance}} \\
\cmidrule(lr{0.5em}){2-5}\cmidrule(lr{0.5em}){6-9}
 & Average Cost & Emission & Price & \,\,\, Grid \,\,\, & Average Cost & Emission & Price & \,\,\, Grid \,\,\, \\
\midrule
RBC  & 0.921 & 0.964 & 0.817 & 0.982 
 & 0.944 & 0.994 & 0.840 & 0.997 \\
MPC & 0.861	& 0.921	& 0.746	& 0.916 
 & 0.906	& 0.965	& 0.820 & 0.933 \\
AMPC  & 0.827	& \underline{0.859} & 0.750 & 0.872
    & 0.901 &	0.914 &	0.835 & 0.955\\
ES  & 0.812	& 0.863	& 0.748	& \underline{0.827}	
& \underline{0.883}	& 0.923 &	0.815 & \underline{0.911} \\
SAC & 0.834 & \underline{0.859} & 0.737 & 0.905 
    & 0.901 & \underline{0.913} & 0.821 & 0.968 \\
PPO & \underline{0.808} & 0.869 & \underline{0.724} & 0.830  
& 0.887 & 0.930 & \underline{0.809} & 0.921 \\
\cmidrule(lr{0.5em}){2-5}\cmidrule(lr{0.5em}){6-9}
\textbf{SOFO} & \textbf{0.795} & \textbf{0.857} & \textbf{0.717} & \textbf{0.810} 
 &  \textbf{0.862} & \textbf{0.911} &	\textbf{0.796} & \textbf{0.881} \\
\cmidrule(lr{0.5em}){2-5}\cmidrule(lr{0.5em}){6-9}
\textit{Improv.} & \textit{+1.3\%} & \textit{+0.2\%} & \textit{+0.7\%} & \textit{+1.7\%} & \textit{+2.1\%} & \textit{+0.2\%} & \textit{+1.3\%} & \textit{+3.0\%} \\
\bottomrule
\end{tabular}
\label{tab:performance}
\end{table*}

\begin{table*}[t]
\centering
% \small
\caption{Effectiveness of the framework components in the testing set. The \textit{Improv.} columns show the relative improvements of each added component, where the positive value indicates that the cost is relatively reduced.}%of SOFO framework over the best performed baselines, respectively.}
\begin{tabular}{lcccccccccc}
\toprule
\multirow{2}{*}{\textbf{}}  & \multicolumn{5}{c}{\textbf{Overall Performance}} & \multicolumn{5}{c}{\textbf{Testing Set Performance}} \\
\cmidrule(lr{0.5em}){2-6}\cmidrule(lr{0.5em}){7-11}
 & Average Cost & \textit{Improv.} & Emission & Price & Grid & Average Cost & \textit{Improv.} & Emission & Price & Grid \\
\midrule
%RBC & 0.965 & 0.971 & 0.831 & 1.094 & 0.928 & 0.944 & 0.753 & 1.088\\
MPC & 0.861	& - & 0.921	& 0.746	& 0.916 
    & 0.906 &-	& 0.965	& 0.820 & 0.933 \\
%	& 0.834 & - & 0.905 & 0.724 & 0.871\\
+ Rolling-Horizon & 0.839 & \textit{+2.2\%} & 0.901 & 0.753 & 0.863 
& 0.896 & \textit{+1.0\%} & 0.946 & 0.812	& 0.929 \\ 
+ Stochastic Optimization 	 & 0.805 & \textit{+3.4\%} & 0.869 & 0.723 & 0.823 
& 0.875 & \textit{+2.1\%} & 0.934 & 0.794 & 0.897 \\
+ Online Fine-tuning & 0.795 & \textit{+1.0\%} & 0.857 & 0.717 & 0.810 
& 0.862 & \textit{+1.3\%}& 0.911 & 0.796 & 0.881\\
%					& 0.788 & & 0.861 & 0.707 & 0.795\\
%PPO & \\	
%MAPPO & 0.856 &	0.891 &	0.759 &	0.920 & 0.807 &	0.872 &	0.700 &	0.850 \\
%\cmidrule(lr{0.5em}){2-5}\cmidrule(lr{0.5em}){6-9}
%\textbf{SOFO} & \textbf{0.808} & \textbf{0.846} &	\textbf{0.728} & \textbf{0.849} & \textbf{0.788} & \textbf{0.861} & \textbf{0.707} & \textbf{0.795} \\
%\cmidrule(lr{0.5em}){2-5}\cmidrule(lr{0.5em}){6-9}
%\textit{Improv} & \textit{1.5\%} & \textit{1.6\%} & \textit{0.7\%} & \textit{2.2\%} & \textit{1.6\%} & \textit{0.7\%} & \textit{0.5\%} & \textit{3.6\%}\\
%\textit{p-value} & \\
\bottomrule
\end{tabular}
\label{tab:components}
\end{table*}

\paragraph{Evaluation}
The evaluation framework follows the official competition setup. Based on the location, the dataset is divided into training set (buildings 1-5), validation set (buildings 6-10), and testing set (buildings 11-17). We focus on the performance on the testing set, as well as the overall performance of the whole district. %This setup follow the task of Citylearn competition, where we aim to provide dispatch solutions for the unseen buildings in the same period. %Two setups are proposed:
%\begin{itemize}
%    \item \textbf{Spatial setup}: Following the setup of NeurIPS Competition, data of 17 buildings are split into training set (building 1-5), validation set (building 6-10) and testing set (building 11-17) according to the location. %This setup follow the task of Citylearn competition, where we aim to provide dispatch solutions for the unseen buildings in the same period. 
%    \item \textbf{Temporal setup}: \textcolor{red}{Alternatively, data are split according to temporal order. Records of the first 8 months are kept for training the model, the following 2 months for validating, and the rest for testing.}
%\end{itemize}
Metrics contain emission cost, price cost and grid cost, which correspond to the goal of economics, decarbonization and fluctuation-reducing, respectively.
Formally, consider there are $I$ buildings and $T$ timestamps. For building $i$ at a timestamp $t$ , denote the load demand as $L_{i,t}$, the solar generation as $P_{i,t}$, the dispatch solution as $X_{i,t}$, the carbon intensity as $c_t$ and the price unit as $p_t$, then we define the electricity consumption of a single building as $E_{i,t} = L_{i,t}-P_{i,t}+X_{i,t}$, and the district-level consumption as $E_{t}^{\rm dist} =\sum_{i=1}^I E_{i,t}$. %Then:
%\begin{equation*}
%\begin{split}
%&\text{Building electricity consumption:}\quad  E_{i,t} = L_{i,t}-P_{i,t}+X_{i,t}\,, \\
%&\text{District electricity consumption:}\quad E_{t}^{\rm dist} =\sum_{i=1}^I E_{i,t}\,.
%\end{split}
%\end{equation*}

With these notations, the evaluation metrics are calculated as: 
%\begin{equation*}
%\begin{split}
%C_{\text {Emission}} = & \sum_{t=1}^T  \left(\sum_{i=1}^I \max \left(L_{i,t}-P_{i,t}+X_{i,t}, 0 \right)\right) \cdot c_t  \,,\\
%C_{\text {Price}} = & \sum_{t=1}^T \max \left(\sum_{i=1}^I\left(L_{i,t}-P_{i,t}+X_{i,t}\right), 0 \right) \cdot p_t \,,\\
%C_{\text {Ramping}}  = & \sum_{t=2}^T  \Bigl(\sum_{i=1}^I  (L_{i,t}-P_{i,t}+X_{i,t}) - 
% \sum_{i=1}^I (L_{i,t-1}-P_{i,t-1}+X_{i,t-1}) \Bigr)\,, \\
%C_{\text {Average}} = &  \left(C_{\text {Emission}} + C_{\text {Price}} + C_{\text {Ramping}} \right) / 3 \,.
%\end{split}
%\end{equation*}
\small
\begin{equation*}
\begin{split}
C_{\text {Emission}} = & \sum_{t=1}^T  \left(\sum_{i=1}^I \max \left( E_{i,t}, 0 \right)\right) \cdot c_t  \,, \quad
C_{\text {Price}} =  \sum_{t=1}^T \max \left(E_{t}^{\rm dist}, 0 \right) \cdot p_t \,,\\
C_{\text{Grid}} = & \frac{1}{2}\left(C_{\text {Ramping}}+C_{\text{Load Factor}}\right)  \\ 
= & \frac{1}{2} \left( \sum_{t=1}^{T-1}  \Bigl|E_{t+1}^{\rm dist} - E_{t}^{\rm dist} \Bigr| + \sum_{m=1}^{\text{\#months}} \frac{\text{avg}_{t \in [\text{month } m]} E^{\rm dist}_t}{\text{max}_{t \in [\text{month } m]}E^{\rm dist}_t}\right) \,.
% C_{\text {Average}} = &  \frac{1}{3}\left(C_{\text {Emission}} + C_{\text {Price}} + C_{\text {Grid}} \right)  \,.
\end{split}
\end{equation*}
\normalsize

\paragraph{Baselines}
We evaluate SOFO by comparing with the baselines:
\begin{itemize}
	\item \textbf{RBC}: Rule-based control. We consider a time-dependent strategy, which charges at 10\% of battery capacity from 10am to 2pm, then discharges at equal amount from 4pm to 8pm.
    \item \textbf{MPC} \cite{sultana2017review}: Conventional approach of model predictive control. A simple yet efficient approach of GBDT model \cite{10.5555/3294996.3295074} is considered as the prediction model, followed by a deterministic optimization problem of day-ahead dispatch.
%    \item \textbf{RHC}: A typical variant of MPC with rolling-horizon control.
%    \item \textbf{SMPC} \cite{yu2019uncertainties}: Stochastic extension of MPC, which takes the uncertainty into account.
%    \item \textbf{PPO} \cite{pmlr-v115-wang20b}: Proximal policy optimization, which is a policy gradient method for reinforcement learning.
%    \item \textbf{MAPPO} \cite{yu2021surprising}: A multi-agent variant of PPO algorithm, with higher sample efficiency and wall-clock runtime efficiency.
\end{itemize}
In addition, we compare it with the state-of-the-art methods, which are implemented by top-ranked teams in the competition. Both optimization and reinforcement learning-based approaches are evaluated to provide a comprehensive comparison.
\begin{itemize}
\item \textbf{AMPC} \cite{sultana2017review}: An adaptive version of MPC, with an online update scheme of forecasting models. 
\item \textbf{ES} \cite{varelas2018comparative}: Evolution strategy with the covariance matrix adaptation for optimization problem.
\item \textbf{SAC} \cite{sac}: Soft Actor-Critic algorithm which controls each agent in a decentralized way with shared parameters.
\item \textbf{PPO} \cite{yu2021surprising}: Proximal policy optimization, which is a recent advancement of reinforcement learning algorithm.
\end{itemize}

\begin{table*}[t]
\centering
% \small
\caption{Comparing the dispatch performance, forecast performance and execution time of SOFO framework with the variation of forecasting models, as well as the approaches of online updating.} 
\begin{tabular}{lccccccccc}
\toprule
\multirow{2}{*}{\shortstack{\textbf{Forecast}\\ \textbf{Model}}} & \multirow{2}{*}{\shortstack{\textbf{Online}\\ \textbf{Update}}}  & \multicolumn{5}{c}{\textbf{Dispatch Performance}} & \multicolumn{2}{c}{\textbf{Forecast Performance}} \\
\cmidrule(lr{0.5em}){3-7}\cmidrule(lr{0.5em}){8-9}
& & Average & Emission & Price & Grid & Time (24h dispatch) & WMAPE-Load & WMAPE-Solar \\
\midrule
\multirow{1}{*}{Linear} &  \multirow{4}{*}{\XSolidBrush} & 
0.878 & 0.929 & 0.806 & 0.899 & \textit{7.96s} & 42.12\%	 & 27.25\% \\
\multirow{1}{*}{GBDT} & &	0.875 & 0.934 & 0.794 & 0.897 & \textit{8.17s} & 44.70\% & 10.74\% \\
\multirow{1}{*}{RNN} & & 0.876 & 0.921 & 0.805 & 0.902 & \textit{9.30s} & 45.97\% & 10.66\%  \\
\multirow{1}{*}{Transformer} &  & 0.879 & 0.920 & 0.802 & 0.916 & \textit{10.64s} &  45.25\% & 10.60\% \\ 
 \cmidrule(lr{0.5em}){1-2} \cmidrule(lr{0.5em}){3-7}\cmidrule(lr{0.5em}){8-9}
\multirow{1}{*}{Linear} &  \multirow{4}{*}{\shortstack{\Checkmark \\ Self-Adaptive \\ Linear Correction}} & 0.871 & 0.918 & 0.804 & 0.890 & \textit{8.17s} & 39.35\% & 21.23\% \\
\multirow{1}{*}{GBDT} & & 0.868 & 0.913 & 0.801 & 0.889 & \textit{8.99s} & 39.48\% & 9.38\%\\
\multirow{1}{*}{RNN} & & 0.866 & 0.913 & 0.802 & 0.888 & \textit{10.01s} & 39.29\% & 9.25\%\\
\multirow{1}{*}{Transformer} & & 0.869 & 0.913 & 0.802 & 0.892 & \textit{11.03s} & 39.86\% & 9.12\% \\ 
% \cmidrule(lr{0.5em}){1-2} \cmidrule(lr{0.5em}){3-7}\cmidrule(lr{0.5em}){8-9}
%\multirow{1}{*}{Linear} &  \multirow{4}{*}{\shortstack{\Checkmark \\ Repeat Learning \\ from Scratch}}\\
%\multirow{1}{*}{GBDT} & & \\
%\multirow{1}{*}{RNN} & & \\
%\multirow{1}{*}{Transformer} & &  \\ 
 \cmidrule(lr{0.5em}){1-2} \cmidrule(lr{0.5em}){3-7}\cmidrule(lr{0.5em}){8-9}
\multirow{1}{*}{RNN}  &  \multirow{2}{*}{\shortstack{\Checkmark \\ Online Fine-tuning}} 
& \textbf{0.862} & \textbf{0.911} & \textbf{0.795} & \textbf{0.881} & \textit{11.45s} & \textbf{38.98\%} & \textbf{9.01\%} \\
\multirow{1}{*}{Transformer} & & 0.864 & 0.912 & 0.799 & 0.880 & \textit{12.15s} & 39.30\% &9.07\% \\ 
%\multirow{2}{*}{GBDT} & \XSolidBrush 
%\multirow{2}{*}{Linear} & \XSolidBrush \\
%& \Checkmark \\
%\cmidrule(lr{0.5em}){3-7}\cmidrule(lr{0.5em}){8-12}
%\multirow{2}{*}{GBDT} & \XSolidBrush \\
%& \Checkmark \\
%\cmidrule(lr{0.5em}){3-7}\cmidrule(lr{0.5em}){8-12}
%\multirow{2}{*}{RNN} & \XSolidBrush \\
%& \Checkmark \\
%\cmidrule(lr{0.5em}){3-7}\cmidrule(lr{0.5em}){8-12}
%\multirow{2}{*}{Transformer}  & \XSolidBrush \\
%& \Checkmark \\
%\cmidrule(lr{0.5em}){3-7}\cmidrule(lr{0.5em}){8-12} 
\bottomrule
\end{tabular}
% The \textit{Improv} and \textit{p-value} rows show the relative improvements and the statistical significance of SOFO framework over the best performed baselines, respectively.}
\label{tab:abl_forecast}
\end{table*}

\paragraph{Implementation}
The environment simulator and evaluation framework are implemented with CityLearn \cite{vazquez2019citylearn}. 
The deep neural network is implemented using PyTorch.
The optimization problem is solved using MindOpt \cite{mindopt}. We run all the experiments on a machine
equipped with Intel(R) Xeon(R) Platinum 8163 CPU @ 2.50GHz, and Nvidia Tesla v100 GPU.

\subsection{Performance Comparison}
Table \ref{tab:performance} illustrates the performance of the proposed method.
All cost presented above are normalized against the simple baseline without electrical energy storage in batteries, such that lower values in the table are preferred. Relative improvements of SOFO over the best-performed baseline are also reported.
In particular, the proposed SOFO framework has improved over the state-of-the-art baselines in all metrics, as indicated by the bold scores in Table \ref{tab:performance}. For example, SOFO contributes to reduce 2.1\% average cost,  0.2\% carbon emission, 1.3\% price cost and 3.0\% grid fluctuation, compared with the best-performed baselines for each metric on the testing set.
% \textcolor{red}{TODO: Adding more interpretation?}
For further investigation, we analyze how the essential model components impact the model performance. As shown in Table \ref{tab:components}, the modules of rolling-horizon control, stochastic optimization and online fine-tuning are added in succession on the MPC method. We observe that all modules can significantly impact performance, which shows the indispensibility of these modules.

% Ablation experiments are conducted over several hyperparameters to understand their impact on SOFO framework.

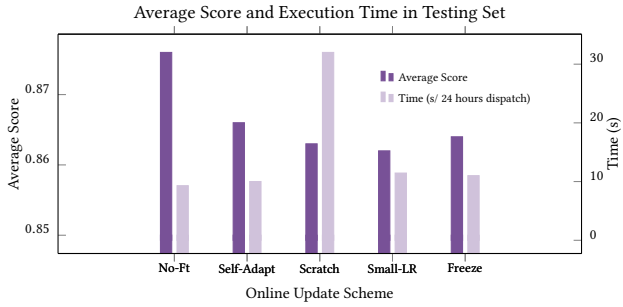
\begin{figure}
\centering 
\begin{tikzpicture}
\begin{axis}[barstyle,width=0.48\textwidth, height=4.5cm, title={}, xlabel={}, ylabel={Average Score}, bar width=1.5mm, symbolic x coords = {No-Ft, Self-Adapt, Scratch, Small-LR, Freeze}, xtick = data, legend style={draw=none, at={(0.4,0.8)}, enlarge x limits=0.4, anchor=west, nodes={scale=0.9, transform shape}, legend image post style={scale=0.5}}, axis y line*=left]
\addplot[color=mycolor2, fill=mycolor2] coordinates { 
            (No-Ft, 0.876)
            (Self-Adapt, 0.866)
            (Scratch, 0.863)
            (Small-LR, 0.862)
            (Freeze, 0.864)
            };
\addplot [color=mycolor2!35, fill=mycolor2!35] coordinates { 
            (No-Ft, 0.85)
            (Self-Adapt, 0.85)
            (Scratch, 0.85)
            (Small-LR, 0.85)
            (Freeze, 0.85)
};
%\legend{Average Score, Time / Day}
\end{axis}
\begin{axis}[barstyle,width=0.48\textwidth, height=4.5cm, title={Average Score and Execution Time in Testing Set}, xlabel={Online Update Scheme}, ylabel={Time (s)}, bar width=1.5mm, symbolic x coords = {No-Ft, Self-Adapt, Scratch, Small-LR, Freeze}, xtick = data, legend style={draw=none, at={(0.6,0.75)}, enlarge x limits=0.4, anchor=west, nodes={scale=0.9, transform shape}, legend image post style={scale=0.5}}, axis y line*=right, legend columns=1, legend cell align={left}]
\addplot[color=mycolor2, fill=mycolor2] coordinates { 
            (No-Ft, 0.876)
            (Self-Adapt, 0.866)
            (Scratch, 0.863)
            (Small-LR, 0.862)
            (Freeze, 0.864)
     	};
\addplot [color=mycolor2!35, fill=mycolor2!35] coordinates { 
            (No-Ft, 9.30)
            (Self-Adapt, 10.01)
            (Scratch, 32)
            (Small-LR, 11.45)
            (Freeze, 11.01)
};
\legend{Average Score, Time (s/ 24 hours dispatch)}
\end{axis}
\end{tikzpicture}
\caption{Evaluation of average score and execution time with respect to different online parameter update schemes.}
        \label{fig:ablation}
\end{figure}

\subsection{Ablation Study}\label{sec:abl}
Ablation experiments are conducted over several modules and hyperparameters to understand their impact on SOFO framework.
\paragraph{Effectiveness of Online Data Augmentation}
We compare the performance of various online update approaches as shown in Figure \ref{fig:ablation}: (1) No-Ft: Without using fine-tuning for online data; (2) Self-Adapt: Self-adaptive linear correction by minimizing the MSE between  the historical real values and predicted values; (3) Scratch: Learning from scratch repeatedly; (4) Small-LR: Continual learning with online data, but using a smaller learning rate; (5) Freeze: Continual learning with online data, but freezing the weights of the first few layers and keeping only the last layer updated.
To compare the model efficiency, we evaluate the average execution time of real-time dispatch in a period of 24 hours.
Empirical results demonstrate the advantages of fine-tuning with a smaller learning rate, considering both efficiency and effectiveness.
% \textcolor{red}{
% Note that we compare the online fine-tuning scheme with the self-adaptive linear correction, which is the original winner solution of CityLearn Challenge 2022.}
%It compares the historical real values (\textit{e.g.}, ${L}_{u,t}$) and the predicted values (\textit{e.g.}, $\hat{L}_{u,t}$). %The self-adaptive correction parameter is updated at each timestep $T_{\rm sa}$ as:
% $\underset{\beta}{\operatorname{argmin}} \sum_{t=1}^{T_{\rm sa}}\left(\beta \cdot \hat{L}_{u,t}-L_{u,t}\right)^2 = \frac{\sum_{t=1}^{T_{\rm sa}} L_{u,t} \cdot \hat{L}_{u,t}}{\sum_{t=1}^{T_{\rm sa}} {\hat{L}_{u,t}}^2} \,.
% $
% It compares the historical real values and the predicted values (\textit{e.g.}, to calculate a linear correction parameter.

% fine-tuning techniques:
% \begin{itemize}
%     \item No-Ft: Without using fine-tuning for online data;
%     \item Self-Adapt: Self-adaptive linear correction.
%     \item Scratch: Learning from scratch repeatedly.
%     \item Small-LR: Continual learning with online data, but using a smaller learning rate;
%     \item Freeze: Continual learning with online data, but freezing the weights of the first a few layers.
% \end{itemize}

\paragraph{Effectiveness of Forecasting Models}
As demonstrated in Table \ref{tab:abl_forecast}, different forecasting models are compared in our framework. Apart from dispatch performance and execution time, the forecasting performance is also reported with weighted mean absolute percentage error (WMAPE). Empirically, the RNN forecasting model with online fine-tuning has achieved superior performance compared to other setups, with low incremental computational cost.

\begin{figure}[t] 
\centering
\begin{tikzpicture}
\begin{axis}[linestyle, width=0.48\textwidth, height=5.2cm, title = Average Score in Testing Set, xlabel={Scenario Number}, legend columns=1,symbolic x coords={1,25,50,75,150,300,450}, legend style={draw=none, at={(0.84,0.76)},anchor=east, nodes={scale=0.9, transform shape}}, legend image post style={scale=0.6}]
\addplot[mark=*, color=mycolor2] coordinates {
(1, 0.901 )
(25, 0.882 )
(50, 0.869 )
(75, 0.862 )
(150, 0.861 )
(300, 0.861 )
(450, 0.861 )
};
\addplot [name path=upper,draw=none] coordinates { 
(1, 0.921 )
(25, 0.894 )
(50, 0.877 )
(75, 0.865)
(150, 0.863 )
(300, 0.862 )
(450, 0.861 )
};
\addplot [name path=lower,draw=none]  coordinates { 
(1, 0.881)
(25, 0.870 )
(50, 0.861)
(75, 0.859)
(150, 0.859 )
(300, 0.860 )
(450, 0.861 )
};
\addplot [fill=mycolor2!20] fill between[of=upper and lower];
\end{axis}
\end{tikzpicture}
\caption{Hyperparameter study: different scenario number $N$. The curve represents the expectation, and the area represents the standard deviation over the stochastic samples.}
\label{fig:diff_dim}
\end{figure}
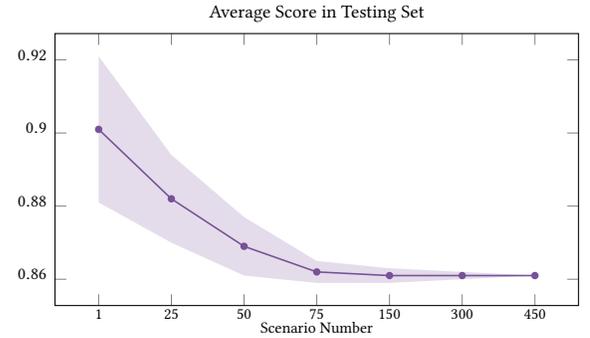

\paragraph{Ablation Study of Stochastic Optimization}
We also analyze the effect of scenario number, as plotted in Figure \ref{fig:diff_dim}. When the scenario number increases, the expectation of the score has converged to fix value around 0.862, with the standard deviation decreasing. %In order to balance the performance and computational complexity, in practice, we choose scenario 

\section{Conclusion}
This paper explores the application of real-time energy dispatch in VPPs under uncertainty. To achieve this, we design an application-driven learning framework to investigate various forms of uncertainties in the control process. %including data drift, model discrepancy and environment perturbation. 
In this framework, two data-driven modules are proposed: a hybrid forecast-and-optimize sequential task integrating deep learning-based forecasting and stochastic optimization, and an efficient online data augmentation scheme involving model pre-training and fine-tuning stages. From empirical point of view, the proposed framework won the 1st place of the CityLearn Challenge 2022, and its effectiveness is further interpreted according to comprehensive experiments in the real-life scenario against various state-of-the-art methods.
%The key insights observed from the empirical results consist of

\section*{Acknowledgment}
We thank the organizers of CityLearn Challenge 2022 who provide an influential competition and high-quality dataset for researchers to investigate the AI application in energy domain. We thank other competition participants who shares pearls of wisdom in the discussions during and after the competition.
We thank Professor Wotao Yin who provided insight and expertise that greatly assisted the research. 
We would also like to show our gratitude to Jiayu Han for technical support during this research.

% \textcolor{red}{In the future,}

% \section*{Ethical Statement}
% There are no ethical issues.

% \section*{Acknowledgments}
% National Key R&D Program of China (2022YFB2403500)

% The preparation of these instructions and the \LaTeX{} and Bib\TeX{}
% files that implement them was supported by Schlumberger Palo Alto
% Research, AT\&T Bell Laboratories, and Morgan Kaufmann Publishers.
% Preparation of the Microsoft Word file was supported by IJCAI.  An
% early version of this document was created by Shirley Jowell and Peter
% F. Patel-Schneider.  It was subsequently modified by Jennifer
% Ballentine, Thomas Dean, Bernhard Nebel, Daniel Pagenstecher,
% Kurt Steinkraus, Toby Walsh, Carles Sierra, Marc Pujol-Gonzalez,
% Francisco Cruz-Mencia and Edith Elkind.

%% The file named.bst is a bibliography style file for BibTeX 0.99c
% \newpage
% \bibliographystyle{named}
% \bibliography{ijcai23}

\balance
\bibliographystyle{ACM-Reference-Format}
\bibliography{ijcai23}

\end{document}